\def\tils{{\tilde s}}

\def\lieg{\eufm g}

\def\tilM{{\hbox{\kern 3pt $\widetilde  {\NoBlackBoxes\hbox 
to 10pt { \kern -3 pt $M$}}$}  }}

\def\tilU{\widetilde U}

\def\tilpi{\tilde \pi}

\def\tilkappa{\tilde \kappa}

\def\tilDelta{\tilde \Delta}
\def\tillambda{\tilde \lambda}

%
%
%
%
%
%
%
%
%

%

\def\State#1 {\statementtag#1 }

\catcode`\^^J=10
\magnification=\mag
\documentstyle{amsppt}
\pagewidth{6.00 truein}
\pageheight{8.00 truein}
\hcorrection{.4 truein}
\vcorrection{.25 truein}

\font\deffont=cmbxti10

\topskip= 28pt
\baselineskip     = 16 true pt  

\TagsOnRight

%

\def\ikedocument{
\ifbibmakemode\immediate\openout1= \jobname.bib\fi
\ifmultisection\immediate\openout2=\jobname.eqn\fi}

%

\headline={
\ifnum\pageno=\lastsectpageno\hfil \else
{\ifodd\pageno\rightheadline \else\leftheadline\fi}\fi
}

\footline={\hss\the\pageno\hss}
\def\pagecontents{\ifvoid\topins\else\unvbox\topins\fi
\dimen0=\dp255 \unvbox255
\ifvoid\footins\else\vskip\skip\footins\footnoterule\unvbox\footins\fi}

%
%
%
%

\newif\ifproofmode    \proofmodefalse
\newif\ifbibmakemode  \bibmakemodefalse
\newif\ifbibcitemode  \bibcitemodefalse
\newif\ifmultisection \multisectionfalse
\newif\ifmulteq        \multeqtrue

%

\newcount\lastsectpageno
\lastsectpageno=0
\define\sectpagenocs#1{sect\expandafter\romannumeral#1pageno}

\define\initializesection#1{\sectionnumber=#1%
\equationnumber=1 \statementnumber=0\exercisenumber=0
\ifmultisection
\global\advance\lastsectpageno by 1
\pageno=\lastsectpageno
\immediate\write2{\noexpand\expandafter\def%
\noexpand\csname\sectpagenocs{\the\sectionnumber}\noexpand\endcsname

{\the\pageno}}
\fi}

\define\forwardsectpageno{\immediate\write2{\lastsectpageno=\the\pageno}}

\define\tocpageno#1{
\expandafter\csname\sectpagenocs{#1}\expandafter\endcsname}

%

\newcount\equationnumber
\newcount\eqnolet
\newcount\sectionnumber
\newcount\statementnumber
\newcount\exercisenumber
\def\strutdepth{\dp\strutbox}

\def\margintagleft#1{\strut\vadjust{\kern-\strutdepth
{\vtop to 
\strutdepth{\baselineskip\strutdepth\vss\llap{\sevenrm
#1\quad}\null}}}}

\def\ifundefined#1{\expandafter\ifx\csname#1\endcsname\relax}%

\def\eqcs#1{s\expandafter\romannumeral\the\sectionnumber%
eq\romannumeral#1}

\def\writeeqcs#1{\ifmultisection%
\immediate\write2{\noexpand\expandafter\def%
\noexpand\csname\eqcs#1\noexpand\endcsname{\the\equationnumber}}\fi%

}

\def\makeeqcs#1{\expandafter\xdef\csname\eqcs#1\endcsname%
{\the\equationnumber}%
}

\def\equationtag#1#2{
\ifundefined{\eqcs#1}
\else
\message{Tag \the\sectionnumber.#1.#2 already exists}
\fi%
\tag"(\the\sectionnumber.\the\equationnumber#2)%
\ifproofmode\rlap{\quad\sevenrm#1}\fi%
\writeeqcs{#1}\makeeqcs{#1}%
\global\eqnolet=\equationnumber%
\global\advance\equationnumber by 1%
\global\def\eqrefnumber{#1}"%
}

\def\Tag#1 {\equationtag{#1}{\empty}}

\define\Tagletter#1 {\tag\the\sectionnumber.\the\eqnolet#1}

\define\equationlabel#1#2#3{%
{\xdef\cmmm{\csname
s\romannumeral#1eq\romannumeral#2\endcsname}\expandafter\ifx
 \csname s\romannumeral#1eq\romannumeral#2\endcsname
\relax\message{Equation #1.#2.#3 not defined 
...}\fi(#1.\cmmm#3)}}

\define\eq#1{\equationlabel{\the\sectionnumber}{#1}{}}

%

\define\statementtag#1 {%
\expandafter\ifx\csname %
s\expandafter\romannumeral\the\sectionnumber %
stat\romannumeral#1\endcsname\relax\else%
\message{Statementtag \the\sectionnumber.#1 already exists 
...}\fi%
\ifproofmode%
\margintagleft{#1}\fi\global\advance\statementnumber by 1%
\expandafter\xdef\csname %
s\expandafter\romannumeral\the\sectionnumber %
 
stat\romannumeral#1\endcsname{\the\statementnumber}\ifmultisection 
\immediate\write2{\noexpand\expandafter\def\noexpand\csname 
s\expandafter\romannumeral\the\sectionnumber%
stat\romannumeral#1%
\noexpand\endcsname{\the\statementnumber}}%
\fi\the\sectionnumber.\the\statementnumber}

\define\statementlabel#1#2{\xdef\cmmm{\csname 
s\romannumeral#1stat\romannumeral#2\endcsname}\expandafter\ifx
 
\csname 
s\romannumeral#1stat\romannumeral#2\endcsname\relax\message{Statement
 
#1.#2 not defined ...}\fi#1.\cmmm}

\define\statement#1{\statementlabel{\the\sectionnumber}{#1}}
\define\st#1{\statementlabel{\the\sectionnumber}{#1}}


\define\exercisetag#1 {%
\expandafter\ifx\csname %
s\expandafter\romannumeral\the\sectionnumber %
exer\romannumeral#1\endcsname\relax\else%
\message{exercisetag \the\sectionnumber.#1 already exists 
...}\fi%
\ifproofmode%
\margintagleft{#1}\fi\global\advance\exercisenumber by 1%
\expandafter\xdef\csname %
s\expandafter\romannumeral\the\sectionnumber %
 
exer\romannumeral#1\endcsname{\the\exercisenumber}\ifmultisection 
\immediate\write2{\noexpand\expandafter\def\noexpand\csname 
s\expandafter\romannumeral\the\sectionnumber%
exer\romannumeral#1%
\noexpand\endcsname{\the\exercisenumber}}%
\fi{\bf\the\sectionnumber.\the\exercisenumber.\quad }}

\define\exerciselabel#1#2{\xdef\cmmm{\csname 
s\romannumeral#1exer\romannumeral#2\endcsname}\expandafter\ifx
 
\csname 
s\romannumeral#1stat\romannumeral#2\endcsname\relax\message{exercise
#1.#2 not defined ...}\fi#1.\cmmm}
\define\Exer#1 {\noindent\exercisetag{#1} }

\def\ex#1{\exerciselabel{\the\sectionnumber}{#1}}
\def\Sol#1{\noindent {\bf 
\exerciselabel{\the\sectionnumber}{#1}\quad}}

%
%
%
%
%

\def\bibyear#1:#2 {\edef\reftag{#1\romannumeral#2}}

\font\slr=cmsl10

\def\cite#1{\catcode`-=11\ifbibmakemode\immediate\write1{#1}[{\bf00}]%

\fi%
\ifbibcitemode%
\bibyear#1  %
\expandafter\ifx\csname\reftag 
bibno\endcsname\relax{\message{#1 not in 
bibfile}[{\bf00}]}%
\else%
[{\slr \csname\reftag bibno\endcsname}\hbox{\kern 1pt}]%
\fi\fi
}
   
\def\referencetag#1#2#3#4#5#6#7#8 
{\edef\reftag{#1#2#3#4#5\romannumeral#6#7#8}
\ifbibmakemode
\immediate\write3{\noexpand\def\csname \reftag 
bibno\endcsname
{\the\refnumb}}
\fi}

\def\referencetag#1:#2 
{\edef\reftag{#1\romannumeral#2}
\ifbibmakemode
\immediate\write3{\noexpand\def\csname \reftag 
bibno\endcsname
{\the\refnumb}}
\fi}

\newcount\refnumb
\refnumb=0
\def\InitializeRef{
\ifbibmakemode\immediate\openout3=\jobname.ref\fi
\NoBlackBoxes
\medskip}

%

\define\today{\ifcase\month\or January \or February \or 
March
\or April  \or May \or June \or July \or August \or 
September 
\or October \or November \or December \fi\space
\oldnos{\number\day}, \oldnos{\number\year}}

\def\hook{\mathbin{\raise2.5pt\hbox{\hbox{{\vbox{\hrule 
height.4pt 
width6pt depth0pt}}}\vrule height3pt width.4pt depth0pt}\,}}

%

\define\Ex#1 {\par\medpagebreak\noindent{\bf Example #1.} }

\define\Rem#1 {\par\medpagebreak\noindent{\bf Remark #1.\ } 
\ }

%

\define\squash#1#2#3{\vspace{-#1\jot}\intertext{#2}\vspace{-#3\jot}}

\define\eqtext#1#2#3{{\vskip -#1\jot}\noindent#2{\vskip 
-#3\jot}}

%

\newcount\endnoteno
\newcount \nb
\endnoteno=1
\nb=1
\def\NB{\ifproofmode${}^{\the\nb}$ \global\advance\nb by 
1\fi}
\def\myitem{\item{\bf[\the\endnoteno]}\global\advance 
\endnoteno by 1} 

%

\def\qed{\hfill\hbox{\vrule width 4pt height 6pt depth 1.5 
pt}}

%
\hyphenation{dif-fer-en-tial di-men-sion-al Helm-holtz
     Cum-mings  Czech-o-sla-vak }

%

\multisectiontrue
\proofmodetrue
\pageno=1

\def\tilDelta{\tilde \Delta}

\catcode`-=11
\def \anderson-felsmcmxcviicbibno {1}
\def \anderson-felsmcmxcixabibno {2}
\def \anderson-fels-torremcmxcixabibno {3}
\def \anderson-fels-torremcmxcixbbibno {4}
\def \bluman-kumeimcmlxxxixabibno {5}
\def \bruning-heintzemcmlxxixabibno {6}
\def \fels-olvermcmxcviiabibno {7}

\catcode`-=12

\proofmodefalse

\bibcitemodetrue

\loadeusm
\loadeufm
\def\reals{{\text {\bf R}}}
\def\bx{{\text {\bf x}}}
\def\bff{{\text {\bf f}}}

\def\Fm{ \alpha }
\def\bq{{\text {\bf q}}}
\def\q{{\text {\bf q}}}
\def\tilM{{\widetilde{M}}}
\def\tilrho{{\tilde{\rho}}}
\def\Inv{{\text{\rm Inv}}}

\ikedocument

\def\rightheadline{}
\vskip 1 in
\heading
Group Invariant Solutions without Transversality and the 
Principle of Symmetric Criticality 
\endheading

\footnote""{\today}
\footnote""{ Research supported  by NSF grants DMS--9804833 
and PHY--9732636}
\vskip 40pt
\centerline{
\vbox{\hsize 130pt
\centerline{  {\smc Ian M. Anderson}}
\centerline{Department of Mathematics}
\centerline{Utah State University}
\centerline{Logan, Utah 84322}}\qquad
\vbox{\hsize 130pt
\centerline{ { \smc Mark E. Fels}}
\centerline{Department of Mathematics}
\centerline{Utah State University}
\centerline{Logan, Utah 84322}}\qquad
\vbox{\hsize 130pt
\centerline{ { \smc Charles  G.  Torre}}
\centerline{Department of Physics}
\centerline{Utah State University}
\centerline{Logan, Utah 84322}
}
}
\vskip40pt
\subheading{Abstract} 

We extend Lie's classical method for finding group 
invariant solutions 
to the case of non-transverse group actions.  For this 
extension of Lie's method we identify a local obstruction 
to the principle of symmetric criticality. Two examples of 
non-transverse symmetry reductions for the potential form 
of Maxwell's equations are then examined.

\subheading{KeyWords} Group invariant solutions, symmetric 
criticality.

\newpage
\ikedocument
\pageno=1
\initializesection{1}

\noindent
\head Introduction \endhead

Let $\Delta = 0 $ be a system of differential equations  
and let $\tilDelta = 0 $ be
a second system which is related to the first by a 
geometric transformation. For example, the second system 
could be derived {}from the first by the process of group 
reduction, a  Backlund transformation, or a differential 
substitution. 
The question then arises whether formal geometric 
properties of the second system $\tilDelta = 0 $ can be 
inferred {}from those of the original.  For example, if 
$\Delta = 0 $ is a system of Euler-Lagrange equations is 
$\tilDelta = 0 $ also a system of Euler-Lagrange equations? 
To answer such questions a rigorous mathematical 
description of the process relating the two systems of 
equations is needed.
In this article we will address this issue in the case 
where $\tilDelta = 0 $ are
the reduced equations for the group invariant solutions to 
$\Delta = 0$.

 Lie's  method of symmetry reduction  for  finding the 
group invariant solutions to partial differential equations 
is well-known (see, for example, Bluman and Kumi 
\cite{bluman-kumei:1989a},  Olver \cite{olver:1993a}, 
Winternitz 
\cite{winternitz:1990a},\cite{winternitz:1992a}). However, 
these references make the hypothesis of {\it 
transversality} of the group action, an assumption which is 
not valid for most problems in field theory and 
differential geometry.  In the next section we show how one 
can dispense with 
this assumption and find the reduced differential equations 
for group actions which are not transverse. 
In section 3 we extend the results in the article 
\cite{anderson-fels:1997c} and find another local 
obstruction to the principle of symmetric criticality 
\cite{palais:1979a} for non-transverse group actions.
Finally in section 4 we find the reduced equations for two 
non-transverse symmetry groups of
Maxwell's equations and we check the principle of symmetric 
criticality on these.

\head 1. Group Invariant Solutions Without Transversality 
\endhead

\subhead 1.1 Preliminaries \endsubhead

 Let  $M$ be an $n$-dimensional  manifold and $\pi\:E\to M$ 
a bundle over $M$.  For this article it is sufficient to  
consider $E$  as a trivial bundle $E= U\times \reals^m$ 
where $U$ is an open set in $\reals^n$.  The manifold $M$ 
will serve as the space of  independent
variables and the bundle $E$  plays the role  of  the total 
space  of  independent and dependent variables.  Points of 
$M$ will be labeled with local coordinates  $(x^i)$ and  
points of  $E$ with local coordinates $(x^i, u^\alpha)$. In 
terms of these coordinates the  projection map $\pi$ is 
given  by  $\pi(x^i,u^\alpha) = (x^i)$. We let $E_x 
=\pi^{-1}(x)$ for $x \in M$.

Let $G$ be a finite  dimensional  Lie group which acts  
smoothly and projectably on
$E$.  That is, the action  of each  element of $G$ is a 
fiber
preserving transformation on $E$ and,  consequently, there 
is a  smooth induced action of $G$ on $M$ such that the 
diagram
$$\CD
       E                    @>g >> E \\
       @V\pi   VV  @VV\pi V   \\
       M           @> g >>     M
\endCD
$$
commutes.  For projectable group actions the action of $G$ 
on the  space of sections of $E$ is given by 
$$
(g\cdot s)(x) =   g \cdot s(g^{-1} \cdot x),
$$
where $s:M \to E$ is a smooth section.

\proclaim{Definition 1.1} Let $G$ be a smooth  projectable 
group action on the bundle
$\pi\:E\to M$ and let $U \subset M$ be open. Then a smooth 
section $s\:U \to E$ is
$G$ {\deffont  invariant},  if for all  $x\in U$ and $g\in 
G$ such that $g\cdot x\in U$,
$$
s(g\cdot x) = g\cdot s(x).
\Tag101
$$
\endproclaim

Let $\Gamma$ be the Lie algebra  of  infinitesimal 
generators for the action of $G$  on $E$. Since the action 
of $G$ on $E$ is assumed projectable
any  basis $V_a$, $a =1$,\dots, $p$, for $\Gamma$ assumes 
the local coordinate form 
$$
V_a = \xi ^i_a(x) \partial _{x^i} + \eta^\alpha_a(x, u) 
\partial _{u^\alpha}   \ .
\Tag103
$$
Suppose that $v_a \in \lieg$ (the Lie algebra of $G$), and 
that $V_a$ is the
corresponding infinitesimal generator. Let $s$ be an 
invariant section given in local coordinates
by $s(x) = (x^i, s^\alpha(x^i))$. If we substitute $g = 
exp(t v_a)$ into equation \eq{101}, 
differentiate with respect to $t$ and put $t=0$ we obtain
the {\deffont infinitesimal invariance equations}
$$
\xi ^i_a(x) \frac{ \partial s^\alpha}{\partial x^i} = 
\eta^\alpha_a(x, s^\alpha) 
\Tag{102}
$$
for the invariant section $s$. If $s$ is  globally defined 
on all of $M$
and if $G$ is connected, then the infinitesimal  invariance 
criterion \eq{102} implies  \eq{101}.

The condition that a group with infinitesimal generators $ 
V_a $ in \eq{103} acts (infinitesimally) transversally is 
(see \cite{bluman-kumei:1989a}, \cite{olver:1993a} ) .
$$
{\text {\rm rank}} [ \xi_1(x),\dots , \xi_p(x) ] = 
{\text {\rm rank}} [ \xi_1(x),\dots , \xi_p(x) , 
\eta_1(x,u), \dots , \eta_p(x,u) ] \ .
$$
Under the assumption of transversality the invariant 
section equation \eq{102} is easily solved
in terms of the infinitesimal invariants for $\Gamma$. 
In the next section we describe how to determine the 
invariant sections
without this hypothesis.

\subhead 1.2 Kinematic Reduction \endsubhead

In order to study group invariant solutions to a 
differential equation we need to
first characterize the $G$ invariant sections of $E$ and 
for this we make the following observation.   Suppose that 
$p\in E$ and  that there is a $G$ invariant section  
$s:U\to E$ with $s(x) = p$, where $x\in U$.  Let
$G_x= \{\,g\in G \,|\, g\cdot x =x \,\}$ be the  {\deffont 
isotropy subgroup of $G$ at $x$.} Then for every  $g\in 
G_x$,  we compute {}from \eq{101}
$$
g\cdot  p = g \cdot s(x) = s( g \cdot x) = s(x) = p.
\Tag121
$$
Consequently, if $g \in G_x$ and $p \in E_x$ and $g\cdot p 
\neq p$, then there can be {\bf no} invariant section 
through the point $p$. In view of this, we
define the {\deffont  kinematic bundle  $\kappa(E)$  for 
the action of $G$ on $E$ } by
$$
\gather
        \kappa(E)  = \bigcup_{x\in M}\kappa_{x}(E) ,
\\
\squash{3}{where}{3}
              \kappa_{x}(E) =\{\, p\in E_x \, |\, g \cdot p 
= p \quad \text{for all}\quad g \in G_x
\,\} .
\endgather
$$
It is easy to  check that  $\kappa(E)$  is a $G$ invariant 
subset of  $E$ and therefore the action of  $G$ restricts 
to an  action on  $\kappa(E)$. The set $\kappa(E)$ is a 
generalization of the construction in 
\cite{fels-olver:1997a}.

Now construct the   quotient  spaces $\tilM= M/G$ and  
$\tilkappa(E) =\kappa(E)/G$  for the actions of $G$ on $M$ 
and  $\kappa(E)$ and define the {\deffont kinematic  
reduction  diagram   for the action of $G$ on $E$ }to be 
the commutative diagram
$$\CD
        \tilkappa(E)               @<\q_{\kappa}  <<    
\kappa(E)                    @>\iota>> E
   \\
            @V \tilde \pi 
VV                                           @V\pi   VV
         @VV\pi V   \\
            \tilM      @< \bq_M <<   M           @> id 
>>     M.
\endCD
\Tag122
$$
In this diagram $\iota$ is   the inclusion map  of  the 
kinematic bundle  $\kappa(E)$  into $E$, the maps $\bq_M$ 
and $\q_{\kappa}$  are the  projection  maps to the   
quotient spaces and $\tilde \pi$ is the  surjective  map  
induced by   $\pi$.  

We mention a few simple technical facts about these 
constructions.
\proclaim{Lemma 1.2 }

\noindent
{\rm i]} For all $ p \in \kappa_x(E)$,  $G_p \cong G_x$.

\noindent
{\rm ii]} If $\tilde{ p } \in \tilkappa(E) $ and $x \in M$ 
satisfy $\tilpi(\tilde p) = \q_M(x)$, then there exists a 
unique point $p \in \kappa_x(E)$ such that $\q_\kappa (p) = 
\tilde {p}$.

\endproclaim

{}from ii] it can be immediately inferred that for every 
local section $\tilde { s}: \tilU \to \tilkappa(E) $ there 
exists a uniquely determined $G$ invariant section $s : 
\q_M^{-1}(\tilU) \to \kappa(E)$ such that
$$
\q_\kappa (s(x)) = \tilde {s} ( \q_M (x)) \ .
\Tag123
$$
The kinematic reduction diagram, along with \eq{123}, leads 
to the following 
existence theorem for invariant sections.

\proclaim{Theorem 1.3 } Suppose that $E$
 admits  a kinematic reduction   diagram such that i)  
$\kappa(E)$ is an  embedded sub-bundle of $E$, ii) the 
quotient spaces
$\tilM$ and $\tilkappa(E)$ are smooth manifolds,  and iii) 
$\tilde {\pi}\:\tilkappa(E) \to \tilM$ is a bundle. Let 
$\widetilde U$ be any open set  in $\tilM$  and  let $U= 
\q_M^{-1}(\widetilde U)$. Then \eq{123} defines a 
one-to-one correspondence between the  $G$ invariant  
smooth sections  $s\:U\to E$  and the smooth sections 
$\tilde {s} \:\widetilde U \to \tilkappa(E)$.
\endproclaim
The proof of this theorem can be found in 
\cite{anderson-fels:1999a}.

For generic group actions the transversality condition is 
replaced by 
the hypothesis that $\iota: \kappa(E) \to E$ is an 
embedding. When $G$ is a compact Lie group acting by 
isometries on a Hermitian vector-bundle $E$ this condition 
holds \cite{bruning-heintze:1979a}.

We now characterize infinitesimally invariant sections. If 
the  rank  of the  coefficient matrix $[\xi^i_a(x)]$ is 
$q$,  then  there are  locally defined functions 
$\phi^a_\epsilon(x)$,
where $\epsilon = 1$,\dots, $p-q$, such that  
$\sum^p_{a=1}\phi^a_\epsilon(x) \xi^i_a(x) =0$.  
Consequently, using the infinitesimal  invariant section 
equation \equationlabel{1}{102}{},  
 we find that
$$
        \sum^p_{a=1}\phi^a_\epsilon(x) \left( 
\xi^i_a(x)\frac{\partial s^\alpha}{\partial x^i} 
+ \eta_a^\alpha (x, s^\beta(x)) \right) = 
        \sum^p_{a=1}\phi^a_\epsilon(x^j) \eta_a^\alpha 
(x^j, s^\beta(x^j)) = 0
\Tag212
$$
are the algebraic equations constraining the invariant 
sections. Accordingly we define the {\deffont 
infinitesimal  kinematic bundle} $\kappa^0(E) = 
\bigcup_{x\in M} \kappa^0_{x}(E)$ where
$$
\align
        \kappa^0_{x}(E)
&             =\{\, p=(x^j,u^\beta) \in E_x \,| 
\,\sum^p_{a=1}\phi^a_\epsilon(x^j) \eta_a^\alpha (x^ j, 
u^\beta) = 0 \, \}
\\
\vspace{2\jot}
&             = \{\, p \in E_x \,| \, Z(p) = 0 
\quad\text{for  all} \quad Z\in \Gamma_x \, \} ,
\Tag213
\endalign
$$
and $\Gamma_x = \{ Z \in \Gamma | \pi_*(Z)(x) = 0 \} $. It 
is easy to
see by using infinitesimal methods that $\kappa(E) \subset 
\kappa^0(E) $ and if, for each $x \in M$, $G_x$ is 
connected, then $\kappa(E) = \kappa^0(E)$. 

The (infinitesimal)  kinematic reduction  diagram \eq{122} 
in local coordinates takes the following form
$$
\CD
        (\tilde{x}^r, v^a)              @<\q_{\kappa}  
<<    (\tilde{x}^r, \hat{x}^k, v^a)
         @>\iota>>
            ( \tilde{x}^r, \hat{x}^k, 
\iota^\alpha(\tilde{x}^r, \hat{x}^k, v^a))   \\
            @V \tilde{\pi} VV                              
@V\pi  VV
      @V\pi VV   \\
            (\tilde{x}^r)     @< \q_M<<            
(\tilde{x}^r, \hat{x}^k)    @>id >>    (\tilde {x}^r, 
\hat{x}^k) .
\endCD
\Tag125
$$
To derive this coordinate description of the kinematic 
reduction diagram we begin with the local coordinates  
$\tilde{\pi}\: (\tilde{x}^r, v^a) \to (\tilde{x}^r)$ on the 
bundle $\tilkappa(E) \to \tilM$.  Since
$\q_M\:M\to \tilM$ is a submersion, we can use the  
coordinates $\tilde{x}^r$ as  part of a local coordinate 
system $(\tilde{x}^r, \hat{x}^k)$ on $M$ where $k = 
1..n-q$, and $q$ is the dimension of the orbits of $G$ on 
$M$.  As a consequence of Lemma 1.1 ii] one can  use
$(\tilde{x}^r, \hat{x}^k, v^a) $ as a system of  local 
coordinates on $\kappa(E)$.
Let $(\tilde{x}^r, \hat{x}^k, u^\alpha) \to 
(\tilde{x}^r,\hat{x}^k)$ be a
system of local coordinates on $E$.  Since $\kappa(E)$ is 
an embedded sub-bundle of $E$, the inclusion map 
$\iota\:\kappa(E)\to E$   assumes the form
$$
        \iota(\tilde{x}^r,\hat{x}^k, v^a)
                =(\tilde{x}^r, \hat{x}^k,  
\iota^\alpha(\tilde{x}^r, \hat{x}^k, v^a)).
\Tag124
$$

If $v^a = \tilde{s}^a(\tilde{x}^r)$ is a local section  of 
$\tilkappa(E)$, then the corresponding $G$ invariant 
section of $E$ is given by
$$
        s^\alpha(\tilde{x}^r,\hat{x}^k) = 
\iota^\alpha(\tilde{x}^r, \hat{x}^k, 
\tilde{s}^a(\tilde{x}^r)).
\Tag126
$$
This is the general local form of an invariant section 
without the transversality assumption which we
will use in the examples.

\subhead 1.3 Dynamic Reduction \endsubhead

Let $\pi^k\:J^k(E) \to M$ be the $k$-th order jet bundle 
of  $\pi\:E\to M$. A point  $\sigma = j^k(s)(x)$ in 
$J^k(E)$ represents the  values of a local section $s$ and 
all its derivatives to  order $k$ at the point $x\in M$.  
Since $G$ acts naturally on the space of sections of $E$ by 
\eq{102}  the action of $G$ on $E$ prolongs to an  action 
on $J^k(E)$ by setting
$$
        g\cdot \sigma  = j^k(g\cdot s)(g x) \quad {\text  
{\rm where} } \ \sigma = j^k(s)(g\cdot x)\ .
$$
Now let $\pi:{\Cal D} \to J^k(E)$ be a vector bundle over 
$J^k(E)$ and suppose that the Lie group $G$ acts 
projectably on ${\Cal D}$ in  a manner which covers the 
action of
$G$ on $J^k(E)$.  A {\deffont  differential operator}  is 
a   section  $\Delta :J^k(E) \to {\Cal D}$.
The differential operator  $\Delta$ is $G$ invariant  if 
$$
        g \cdot \Delta(\sigma) = \Delta (g \cdot  \sigma )
$$
for all $ g\in G$ and all points $\sigma \in J^k(E)$. A   
section $s$ of $E$ defined on an open set $U\subset M$ is a 
solution to the differential equations $\Delta = 0$  if
$$
\Delta(j^k(s)(x)) = 0   \quad\text{for all} \quad x\in  U.
$$
Often the action of $G$ on ${\Cal D}$ is naturally defined 
in terms of the action of $G$ on $E$ as in the case of 
Euler-Lagrange operators (see section 2.2).

We want to construct a  bundle $\widetilde{\Cal D}\to 
J^k(\tilkappa(E))$ and  a differential  operator 
$\tilDelta\:J^k(\tilkappa(E)) \to \widetilde { {\Cal D}} $
such that  the correspondence \equationlabel{1}{123}{}  
defines a 1-1 correspondence between the  $G$ invariant 
solutions of $\Delta=0$ and the solutions of $\tilDelta = 
0$. The required  bundle $\widetilde{{\Cal D}}\to 
J^k(\tilkappa(E))$ can not be  constructed  by a direct 
application of the kinematic  reduction  diagram   to 
${\Cal D} \to J^k(E)$.   This difficulty  is circumvented 
by introducing the  {\deffont  bundle of invariant $k$- 
jets}, (Olver \cite{olver:1993a})
$$
\align
        \Inv^k(E)=\{\, \sigma\in
&       J^k(E) \, |\, \sigma =j^k(s)(x_0),
 \\
\vspace{2\jot}
 &      \text{where } \  s \ \text{ is a $G$ invariant 
section defined in a neighborhood of
        $x_0$}\,\}.
\endalign
$$
The fundamental property we need of $\Inv^k(E)$ is that the 
quotient space  $\Inv^k(E)/ G$  coincides  with the  jet 
space $J^k(\tilkappa(E))$. We let ${\Cal D}_{\sssize \Inv} 
\to \Inv^k(E)$  be the restriction of  ${\Cal D}$ to the 
bundle of invariant sections and to this  bundle we now 
apply
our  reduction  procedure to arrive at the {\deffont  
dynamic reduction diagram}
$$
\CD
\tilkappa({\Cal D}{\sssize \Inv}) @<\q<<    \kappa({\Cal 
D}_{\sssize \Inv}) @>\iota>> {\Cal D}_{\sssize \Inv}  
@>\iota^k >> {\Cal D}
\\
@V\tilde \pi VV                 @V\pi 
VV                               @V\pi VV                
@VV \pi V
\\
J^k(\tilkappa(E)) @<\q_{\sssize \Inv}<<  \Inv^k(E) @> Id 
>>\Inv^k(E)    @> \iota ^k >> J^k(E) .
\endCD
\Tag127
$$
Any $G$ invariant differential operator $\Delta\:J^k(E) \to 
{\Cal D}$  restricts to a  $G$ invariant differential 
operator
$\Delta\:\Inv^k(E)\to {\Cal D}_{\sssize \Inv}$ which 
determines a  differential operator 
$\tilDelta\:J^k(\tilkappa(E)) \to \tilkappa({\Cal 
D}_{\sssize \Inv}) $. We call 
the differential operator the reduced operator and the 
equations $\tilDelta = 0 $ the 
reduced equations. It is not difficult to prove the 
following theorem.

\proclaim{Theorem 1.4} The solutions to the reduced 
differential equations  $\tilDelta = 0 $ are in one-to-one 
correspondence with  the $G$ invariant solutions  for  the 
original equations $\Delta=0$.
\endproclaim

To describe diagram \eq{127} in local  coordinates,  we 
begin with the  coordinate  description
 \eq{125} of the kinematic reduction diagram and we let  
$$
        (\tilde{x}^r, \hat{x}^k, u^\alpha, u^\alpha_r, 
u^\alpha_k, u^\alpha_{rs}, u^\alpha_{rk}, u^\alpha_ 
{kl},\ldots)
\Tag199
$$
be coordinates on $J^k(E)$. Since the invariant sections   
are  parameterized by  functions $v^a= v^a(\tilde{x}^r)$ 
the  coordinates for  $\Inv^k(E)$ are
$$
        (\tilde{x}^r, \hat{x}^k, v^a, v^a_r, v^a_{rs},\dots)
$$
and the inclusion map
$$
        \iota^k \:\Inv^k(E) \to J^k(E)
$$
 is  given by
$$
        \iota^k(\tilde{x}^r, \hat{x}^k, v^a, v^a_r, 
v^a_{rs},\ldots) =
        (\tilde{x}^r, \hat{x}^k, u^\alpha, u^\alpha_r, 
u^\alpha_k, u^\alpha_{rs}, u^\alpha_{rk}, u^\alpha_{kl}, 
\ldots),
\Tag128
$$
where the derivative terms are obtained by taking 
derivatives of $u^\alpha = \iota^\alpha(\tilde{x}^r, 
\hat{x}^k, v^a)$, for example,
$$
   u^\alpha_r =         \frac{\partial 
\iota^\alpha}{\partial\tilde{x}^r} + 
\frac{\partial\iota^\alpha}{\partial v^a} v^a_r,
\qquad
               u^\alpha_k
=           \frac{\partial \iota^\alpha}{\partial 
\hat{x}^k} .
$$
The coordinates on $J^k(\tilkappa(E))$ are then
$$
(\tilde{x}^r, v^a, v^a_r, v^a_{rs},\ldots).
$$

Next let $\bff^A$  be a local frame  field for the vector 
bundle ${\Cal D}$.   The  differential operator  
$\Delta\:J^k(E) \to {\Cal D}$ can be written
 in terms of the  coordinates \eq{199} on $J^k(E)$ and this 
local frame as
$$
        \Delta=
        \Delta_A( \tilde{x}^r, \hat{x}^k, u^\alpha, 
u^\alpha_r, u^\alpha_k, u^\alpha_{rs}, u^\alpha_{rk}, 
u^\alpha _ {kl} \ldots)\, \bff^A.
\Tag129
$$
The restriction of $\Delta$ to $\Inv^k(E)$ defines  the 
section  $\Delta_{\Inv}\:\Inv^{k}(E) \to {\Cal D}_{\sssize 
\Inv}$
by
$$
        \Delta_{\Inv} = \Delta_{\Inv,A}(\tilde{x}^r, 
\hat{x}^k , v^a,         v^a_r, v^a_{rs},\ldots) \, \bff^A,
\Tag1210
$$
where the component functions are 
$\Delta_{\Inv,A}(\tilde{x}^r, \hat{x}^k, , v^a, v^a_r, 
v^a_{rs},\ldots)= \Delta_A \circ \iota ^k $.   Since 
$\Delta$  is a $G$ invariant differential
operator, $\Delta_{\Inv}$  is a $G$ invariant differential 
operator and  hence  necessarily factors through the 
kinematic bundle $\kappa({\Cal D}_\Inv)$, and is a section
$$
        \Delta_{\Inv}\:\Inv^k(E) \to \kappa({\Cal 
D}_{\sssize \Inv}) \ .
$$
The existence theorem for invariant sections  implies that 
we  can also find a locally defined   $G$ invariant frame  
$\bff^Q_\Inv$  for $\kappa({\Cal D}_{\sssize \Inv})\to 
\Inv^k(E)$ in terms of which the invariant operator  
$\Delta_{\Inv}$ can be  expressed as
$$
        \Delta_{\Inv}=
               \Delta_{\Inv,Q}( \tilde{x}^r, \hat{x}^k, 
v^a, v^a_r, v^a_{rs},\ldots) \, \bff^Q_\Inv.
$$
The reduced operator is now easily determined.

\bigskip
\bigskip

\sectionnumber=2
\equationnumber=1
\statementnumber=0

\head 2.  The Principle of Symmetric Criticality \endhead

\subhead 2.1 Lagrangian Reduction \endsubhead

Let $\lambda = L \nu $ be a $G$ invariant $k^{th}$ order 
Lagrangian, where $L$ is a smooth function on $J^k(E)$ and 
$ \nu $ is a volume form on $M$. For simplicity $\nu$ is 
assumed to be $G$ invariant and $L$ a differential 
invariant.  
In order to define the reduced Lagrangian $\tillambda$ on 
$J^k(\tilkappa(E))$, we start with 
 a $G$ invariant $q$ chain $\chi : M \to \wedge^q(TM)$ and 
define the 
equivariant bundle map
$$
\rho_\chi : \wedge^*(T^* M) \to \wedge^{*-q} (T^* M),
$$
by $\rho_\chi(\alpha) = \chi \hook \alpha $. The map 
$\rho_\chi$ induces a map on $G$ invariant sections 
$$ 
\rho_\chi : \Omega^*(M)^G \to \Omega^{*-q}(M)^G\ . 
$$
The image of a differential form under $\rho_\chi$ is a $G$ 
basic form, {}from which we further obtain a map 
$$
{\tilde \rho}_\chi: \Omega^{*}(M)^G \to \Omega^{*-q}(\tilM) 
\ .
$$
For example, if $G=SO(3)$ is acting in the standard way on 
$M=\reals^3-(0,0,0)$, then (\cite{anderson-fels:1997c} pg. 
615)
$$
\chi = r \left( z \partial _x \wedge \partial _y - y 
\partial _x\wedge \partial _z + x \partial _y \wedge 
\partial _z \right),
\Tag200
$$
where $r = \sqrt{x^2+y^2+z^2}$, and 
$$
\rho_\chi( dx \wedge dy \wedge dz) = r (x\, dx + y\, dy + 
z\, dz)  
$$
so 
$$
{\tilde \rho}_\chi( dx \wedge dy \wedge dz) = r^2 dr  \ .
$$
Again see \cite{anderson-fels:1997c} or 
\cite{anderson-fels-torre:1999b} for more details and 
examples. 

Using the maps ${\tilde \rho}_\chi$ and $\iota^k :\Inv ^k 
(E) \to J^k(E)$ we define the reduced Lagrangian as
$$
\tillambda = {\tilde \rho}_\chi( (\iota^k)^* \lambda ) \, .
\Tag209
$$
We point out that the role of the map ${\tilde \rho}_\chi$ 
is to reduce the number of independent variables in the 
volume form.

\smallskip
Let $Vert(E)$ be the vertical bundle for $\pi:E \to M$. 
Then {}from the maps $\pi^{2k}_E:J^{2k}(E) \to E $ and 
$\pi^{2k}_M :  J^{2k}(E) \to M $ we construct the {\deffont 
bundle of source forms} ${\Cal D} \to J^{2k}(E)$ by
$$
{\Cal D} = (\pi^{2k}_E)^*\left( Vert^*(E) \right) \wedge 
(\pi^{2k}_M)^* \left( \wedge^n T^*M \right)
$$
where ${\ }^*$ denotes pullback. A point $\omega \in {\Cal 
D}_\sigma, \ \sigma \in J^{2k}(E) $ is a differential form
$$
\omega = A_\alpha \, du^\alpha \wedge \nu \ .
$$
The Euler-Lagrange form ${\Cal E}(\lambda) $ of a $k^{th}$ 
order Lagrangian is a section of ${\Cal D}$ and may be 
written in local coordinates $(x^i,u^\alpha )$ as
$$ 
{\Cal E}(\lambda) = E_\alpha (L) d u^\alpha \wedge \nu \ ,
$$ 
where $E_\alpha(L)$ are the Euler Lagrange expressions for 
$\lambda = L \nu $. 

At this point there are two ways to proceed in the 
reduction of the Euler-Lagrange equations. First, given a 
$G$ invariant Lagrangian $ \lambda$ the Euler-Lagrange 
operator ${\Cal E}(\lambda)$ is
also $G$-invariant. Using the dynamic reduction diagram 
\equationlabel{1}{127}{} in section 1.3 we can then compute 
the reduced operator $\widetilde {{\Cal E}(\lambda)} $ 
whose solutions determine the group invariant solutions. 
The reduced operator $ \widetilde {{\Cal E}(\lambda)} $ is 
a section of $\tilkappa( {\Cal D}_{\Inv})$. Alternatively, 
if $\tillambda$ is the reduced Lagrangian, then ${\Cal 
E}(\tillambda)$ is a source form on $J^{2k}(\tilkappa(E))$ 
which, using the coordinates in \equationlabel{1}{125}{}
$$
{\Cal E}(\tillambda) = E_a(\tilde L) dv^a \wedge {\tilde 
\rho}_\chi (\nu)  
$$
where  ${\tilde \rho}_{\chi}(\nu) $ is a volume form
 on $\tilM$. Therefore it is {\bf not} possible to compare 
the sections $ \widetilde {{\Cal E}(\lambda)} $ and ${\Cal 
E}(\tillambda) $ since they are sections of different 
bundles and consequently, the solutions to $\widetilde 
{{\Cal E}(\lambda)}=0$ 
and ${\Cal E}(\tillambda) = 0 $ may be different. This 
leads to the following principle.

\proclaim{The Principle of Symmetric Criticality (PSC)} Let 
$G$ be a Lie group acting projectably on $E$ and suppose 
there exists a $G$ invariant chain $\chi$. Let $\lambda$ be 
a fixed but arbitrary
$G$ invariant Lagrangian with $\tillambda$ the reduced 
Lagrangian, and let $ \widetilde {{\Cal E}(\lambda)} $ be 
the reduced Euler-Lagrange operator. If, for every choice 
of $\lambda $ the submanifolds $ {\Cal E}(\tillambda) 
^{-1}(0) \subset J( \tilkappa(E)) $ and $\widetilde {{\Cal 
E}(\lambda)} ^{-1}(0) \subset J(\tilkappa(E))$ coincide, 
then we say the principle of symmetric criticality holds 
for the action of $G$ on $E$.
\endproclaim
Note that if PSC holds then every locally defined solution 
$\tilde s : \tilU \to  \tilkappa(E)$ to ${\Cal 
E}(\tillambda) = 0 $ determines though 
\equationlabel{1}{123}{}   a solution $s$ to ${\Cal 
E}(\lambda)=0$. Of course the principle can often fail, see 
\cite{anderson-fels:1997c}, \cite{palais:1979a}. 

Our objective is to determine the local obstruction to this 
principal which arises for non-transverse group actions. 
\medskip

A local obstruction to the commutation relation
$$
{\Cal E}\left( \tilrho_\chi (\iota^k)^* \lambda \right)  = 
\tilrho_\chi {\Cal E}((\iota^k)^* \lambda) \, .
\Tag201
$$
was found in \cite{anderson-fels:1997c} and is described in 
the following Theorem.
\proclaim{Theorem 2.1} If $H^q(\Gamma_x, G_x) \neq 0 $ for 
all $x\in M$, then there exists
a locally defined map $\tilrho_\chi : \Omega^*(M)^G \to 
\Omega^{*-q}(\tilM)$ 
such that \eq{201} holds, and moreover 
${\tilde \rho}_\chi$ is a co-chain map, i.e. $ d \,{\tilde 
\rho}_\chi ={\tilde \rho}_\chi \, d \ $.
\endproclaim
For a description of the Lie algebra cohomology condition 
$$
H^q(\Gamma_x, G_x) \neq 0 
\Tag{208}
$$
see \cite{anderson-fels:1997c} pg. 650. We now find

\proclaim{Lemma 2.2}  Suppose  that $ H^q(\Gamma, G_x) \neq 
0 $ for all $x \in M$. 
Let $ \tilU \subset \tilM$ and let $\tilde s:\tilU \to 
\tilkappa(E)$ be a solution to the Euler-Lagrange equations 
${\Cal E} (\tillambda) = 0 $ for the reduced Lagrangian. 
Then the corresponding invariant section $s:U  \to E$ {}from 
\equationlabel{1}{123}{} is a solution to $(\iota^{2k})^* 
{\Cal E} ( \lambda) =0 $.
\endproclaim
\demo{Proof} It follows {}from the Lemma 2.1 that
$$
{\Cal E} (\tillambda) = {\Cal E}\left( \rho_\chi 
(\iota^k)^* \lambda \right)  = {\tilde \rho}_\chi {\Cal 
E}((\iota^k)^* \lambda) \, ,
$$
which together with the basic fact about the Euler-Lagrange 
operators \cite{olver:1993a} 
$$
{\Cal E}((\iota^k)^* \lambda) =  (\iota^{2k})^* {\Cal E 
}(\lambda)\ 
\Tag222
$$
proves the lemma. \qed
\enddemo

Note however that this Lemma does {\bf not} guarantee that 
the invariant section $s:U \to E$ is a solution to the 
original Euler-Lagrange equations ${\Cal E}(\lambda ) = 0 
$. Let's see why.

\subhead 2.2 The Palais Condition \endsubhead

In this section we will assume that the cohomology 
condition \eq{208} in Theorem 2.1 holds so
that \eq{201} is valid.  

The obstruction to PSC we have hinted at arises {}from 
differences in the
two types of reductions which are occurring.  First, if 
$\lambda $
is a $G$ invariant Lagrangian, then the Euler-Lagrange 
operator $\Delta = {\Cal E}(\lambda)$ is  $G$ invariant, 
and the right side of the dynamic reduction diagram gives
$$
\Delta_{\Inv} = E_\alpha(L) \big|_{\Inv} du^\alpha \wedge 
\nu  \, ,
$$
and a solution  $\tilde s : \tilU \to \tilkappa(E)$ to the 
reduced equations
$ \widetilde { {\Cal E}(\lambda) } = 0 $ defines through 
\equationlabel{1}{123}{}
an invariant solution $s:U \to E$ to 
$E_\alpha(L)\big|_{j(s)(x)} = 0$.
On the other hand if we reduce the Lagrangian $\lambda$ as 
in \eq{209}, we find, using the formula in \eq{222} 
$$
{\Cal E}\left( (\iota^{k})^* \lambda \right)  = 
(\iota^{2k})^ * {\Cal E}(\lambda) 
= \left( \frac{\partial \iota^\alpha}{\partial v^a} 
E_\alpha(L)\right)\bigg|_{\Inv}  dv^a \wedge \nu \ .
$$
Therefore if $\tilde s :\tilU \to \tilkappa(E)$ is a 
solution to the equations ${\Cal E}(\tillambda) = 0 $
we have by Lemma 2.2 that the corresponding invariant 
section $s$ in \equationlabel{1}{123}{}  is a solution to 
$$
\left( \frac{\partial \iota^\alpha}{\partial v^a} 
E_\alpha(L)\right)\bigg|_{j(s)(x)}  = 0 \ .
$$ 
Clearly this does not guarantee that 
$E_\alpha(L)\big|_{j(s)(x)} = 0 $, or that $\tilde s$ is a
solution to the reduced equations. 

Conditions under which solutions to the equations ${\Cal 
E}(\tillambda) = 0$ are solutions
to $\widetilde { {\Cal E}(\lambda) } = 0 $ can be obtained 
by a detailed analysis of $\Delta_{\Inv}$ and $ 
(\iota^{2k})^*{\Cal E}(\lambda )$. 
\medskip

Suppose ${\Delta} = {\Cal E}(\lambda)$ is an invariant 
section of the bundle of source forms $\Cal D$, and let 
$\sigma \in \Inv^{2k}(E)$ and $x_0 =\pi^M(\sigma) \in M$ 
and $Z \in \Gamma_{x_0}$. Then on account of the invariance 
of $\nu$ the Lie derivative of $\Delta $ is
$$
\left( {\Cal L}_{Z} \Delta \right)\big|_{\sigma}  = \left( 
pr^{2k} Z \, (E_\alpha(L))  + E_\beta (L)\frac{ \partial 
\eta^\beta_a}{\partial u^\alpha} \right)\bigg|_\sigma 
du^\alpha \wedge \nu   .
$$
This leads to the isotropy condition
$$
0  =  \left( \Delta_\beta \frac{ \partial 
\eta^\beta}{\partial u^\alpha} \right)\bigg|_{\sigma} 
du^\alpha \wedge \nu  
\Tag224
$$
which determines the fibres of $\kappa^0({\Cal D}_{\sssize 
\Inv})$. 
A geometric description of \eq{224} can be given in terms 
of the {\deffont vertical linear isotropy representation}. 
Suppose $\kappa(E) \subset E$ is an embedded sub-bundle and 
let $Vert(\kappa(E))$ be the restriction (or pullback) of 
the vertical bundle over $E$ to $\kappa(E)$.  
Since $G$ is fibre-preserving, the differential of the 
action of $G$ on $E$ induces an action of $G$ on $Vert(E)$ 
which, because $\kappa(E) \subset E$ is a $G$ invariant 
set, restricts to an action on $ Vert(\kappa(E))$. We then 
make the following definition.

\proclaim{Definition 2.2} Let $p \in \kappa(E)$,   $g \in 
G_p$ and $Y_p \in Vert_p \kappa(E)$. Then the {\deffont 
vertical linear isotropy representation} on $\kappa(E)$,  
$\rho_p: G_p \to GL(Vert_p \kappa(E))$ is defined by
$$
\rho_p(g) Y_p = g_* Y_p \, . 
\Tag229
$$
\endproclaim
Note by Lemma 1.1 i] this is also a representation of 
$G_x$, where $ x = \pi(p)$.

The differential of $\rho_p$ is a homomorphism 
$\rho_p:\Gamma_{p} \to gl(Vert_p \kappa(E))$ called the
{\deffont infinitesimal vertical linear isotropy 
representation}. If $ Z = \xi^i(x) \partial _{x^i}+ 
\eta^\alpha(x,u) \partial _{u^\alpha} \in \Gamma 
_{(x_0,u_0)}$ then, in the standard basis $\partial 
_{u^\alpha}$, we have
$$
\rho_{(x_0,u_0)}(Z) = \left( \frac{ \partial \eta 
^\beta}{\partial u^\alpha}\bigg|_{(x_0,u_0)}\right) \,  .
$$

Equation \eq{224} now proves the following Lemma.

\proclaim{Lemma 2.4} If $\sigma \in \Inv^{2k}(E)$ and $p = 
\pi_E(\sigma) $,  then $\kappa_{\sigma}({\Cal D}_{\sssize 
\Inv}) \cong [Vert^*_p(\kappa(E)) ]^{G_x} $. 
\endproclaim
The invariants $ [Vert^*_p(\kappa(E)) ]^{G_x} $ are 
computed using the dual of the representation \eq{229}. It 
is not difficult to check

\proclaim{Lemma 2.5} $\iota_* Vert_p(\kappa(E)) = [ 
Vert_{\iota(p)} E ] ^{G_x} $. 
\endproclaim
Define the subspace
$$
\left( [Vert_p(E)]^{G_x} \right) ^\perp = \{ \alpha \in 
Vert_p^*(E) \ |\ \alpha(X) = 0 \ {\text {\rm for \ all} } \ 
X \in  [Vert_p(E)]^{G_x} \} \ .
$$
The pullback construction for ${\Cal D}$ and Lemma 2.5 
allows us to identify 
$\left( [Vert_p(E)]^{G_x} \right) ^\perp$ with the subspace 
of $ {\Cal D}_{\Inv,\sigma} $, where $\pi^{2k}_E(\sigma) = 
p $, given by
$$
\left( [Vert_p(E)]^{G_x} \right)^{\perp} \cong 
\{ A_\alpha du^\alpha \wedge \nu \in {\Cal D}_{\Inv,\sigma} 
\ | \ A_\alpha \frac{\partial \iota^\alpha}{\partial v_a} = 
0 \ \} \, ,
$$
and the coordinates in \equationlabel{1}{128}{} are being 
used.  We now come to the key result.
\proclaim{Theorem 2.5} Suppose that $\lambda $ is a $G$ 
invariant Lagrangian and that
$H^q(\Gamma_x, G_x) \neq 0 $ for all $ x \in M$. If 
$$
[ Vert^*_p(E) ]^{G_x} \cap \left( [Vert_p(E)]^{G_x} \right) 
^\perp = 0  \quad {\text for \ all} \ p \in \kappa(E)\, ,
\Tag223
$$
then every solution $\tils : \tilM \to \tilkappa(E)$ to 
${\Cal E}(\tillambda) =0 $, where $\tillambda $ is reduced 
Lagrangian in \eq{209},  defines by 
\equationlabel{1}{123}{}, a solution to ${\Cal E}(\lambda) 
=0 $. In other words, the PSC holds. 
\endproclaim

We will prove this under the hypothesis that the isotropy 
groups are connected so
that we may use infinitesimal methods.  

\demo{Proof} Let $\tilde {s}: \tilU \to \tilkappa(E)$ be a 
solution to ${\Cal E} (\tillambda) = 0$ and let $s :U\to 
\kappa(E)$ be the corresponding invariant section {}from 
\equationlabel{1}{123}{}. 
 Using the isomorphism in Lemma 2.4, we have
$$
E_\alpha(L)\big|_{j(s)(x)} \in [Vert^*_{(x,s(x))}(E)]^{G_x} 
\ .
$$
We also have, by Lemma 2.2, that $s$ satisfies
$$
\left( \frac{\partial \iota^\alpha}{\partial 
v^a}E_\alpha(L)\right)\bigg|_{j(s)(x)} = 0 ,
$$
therefore by Lemma 2.5 
$$ 
E_\alpha(L)\big|_{j(s)(x)}\in  \left( [Vert_p(E)]^{G_x} 
\right) ^\perp 
$$
where $p = (x,s(x))$. Thus, for each $x \in U$,
$$
E_\alpha(L)\big|_{j(s)(x)} \in [ Vert^*_{(x,s(x))}(E) 
]^{G_x} \cap \left( [Vert_{(x,s(x))}(E)]^{G_x} \right) 
^\perp
$$
which vanishes by \eq{223} and $s$ is an invariant solution 
to $E_\alpha(L)=0$. \qed
\enddemo
It is also possible to prove that \eq{223} is necessary for 
the PSC to hold. We call
\eq{223} {\it the Palais condition} due to its similarity 
to 
the condition given in Palais' original article 
\cite{palais:1979a} .

\bigskip
\bigskip
\sectionnumber=3
\equationnumber=1
\statementnumber=0

\head 3. Example - Maxwell's Equations \endhead

The base manifold in this case is $M = \reals^4 $ with 
coordinates $(t,x,y,z)$ and $E=T^*M$.
The Minkowski metric is $\eta = diag ( +,-,-,- )$ and   
we use $(x^a) , \, a =0,1,2,3$, and $(x,y,z)=(x^i), 
i=1,2,3$ when convenient.
A section of $E$ is a differential form $ \Fm = u_a dx^a $. 
The form $\Fm$ is the potential in Maxwell's equations 
which are the Euler-Lagrange equations of the first-order 
Lagrangian
$$
\lambda = | d \Fm |^2 \nu \ ,
\Tag300
$$ 
where $\nu = dx^0 \wedge dx^1 \wedge dx^2 \wedge dx^3 $ and 
$| d \Fm |^2 = \eta^{ab} \eta^{cd} u_{[a,c]} u_{[b,d]} $. 
The Euler-Lagrange equations are $\Delta = \Delta^a du_a 
\wedge \nu $, where
$$
\align
\Delta^0 & = u_{0,xx}+u_{0,yy}+u_{0,zz} 
-u_{1,tx}-u_{2,ty}-u_{3,tz}\ ,
  \\
\Delta^1 & = -u_{0,tx} +u_{1,tt}-u_{1,yy}-u_{1,zz} 
+u_{2,xy}+u_{3,xz}\ ,\\
\Delta^2 & = -u_{0,ty} + u_{1,xy} 
+u_{2,tt}-u_{2,xx}-u_{2,zz}+u_{3,yz}\ ,  \\
\Delta^3 & = -u_{0,tz} +u_{1,xz} +u_{3,yz} 
+u_{3,tt}-u_{3,xx} -u_{3, yy} .
\endalign
$$

We now consider two non-transverse groups actions, find the 
reduced equations and investigate the principle of 
symmetric criticality.

\proclaim{Example 3.1}  $\Gamma  = \{ \epsilon _{kij} x^i 
\partial _{x^j} -\epsilon _{kij} u_i \partial _{u_j} \} $ 
\endproclaim 

The Lie algebra $\Gamma$ is obtained {}from the infinitesimal 
generators of $SO(3)$ acting in the standard way on $E$. At 
the point $ \bx_0 = (t_0,x^k_0)$ we have the isotropy 
vector-field 
$$
Z = x^k_0 \left( \epsilon _{kij}  x^i\partial_{x^j} - 
\epsilon_{kij} u_i \partial_{u_j} \right) \ \in \Gamma 
_{\bx_0} \ .
\Tag301
$$
The sub-bundle $\kappa(E) \subset E $ at the point $\bx_0$ 
is given by the vanishing of $ x^i_0  \epsilon_{kij} u_i $ 
(the cross product of $(x^i_0)$ and $(u_i)$). Therefore 
$u_0 = v $, $ u_i = w x^i_0 $, with $v,w$ being fibre 
coordinates on $ \kappa_{\bx_0} (E)$. The inclusion $\kappa 
(E) \to E $ is given by
$$
\iota^k (t,x,y,z\, ; \, v, w ) \to (t,x,y,z\, ; \, v, w x 
,w y , w z) \ .
$$
The functions $t$ and $ r =\sqrt{x^2+y^2+z^2}$ are 
invariants on $M$ so that in local coordinates the 
kinematic reduction diagram is
$$
\CD
    (t, r\, ;\, v, w)  @<\q_{\kappa }<< (t,x^i\, ; \, v 
,w)     @> \iota >>  (t,x^i\, ; \, u_a) 
\\
              @V\tilde \pi VV              @V\pi VV      @ 
VV  \pi V
 \\
              (t,r)           @<\q_M < <                
(t,x^i)            @>> id  >  (t,x^i). 
\endCD
\Tag302
$$
The invariant sections are now easily determined to be
$$
u_0 = v(t,r) \ , \ u_i = w(t,r) x^i \ .
$$

In order to determine the reduced equations we compute the 
fibres of ${\Cal D}_{\sssize \Inv}$ by using Lemma 2.4. At 
the point $ p_0 = (t_0,x_0,y_0,z_0\, ; \, v_0, w_0 x_0 ,w_0 
y_0 , w_0 z_0) \in \kappa(E)$  the vector-field $Z$ in 
\eq{301} lies in $\Gamma_{p_0}$. The (infinitesimal) 
vertical linear isotropy representation in the coordinate 
basis $\partial _{u_a}$ is
$$
\rho_{p_0} (Z)  =  \frac{ \partial \ }{ \partial u_l}   
x_0^k \epsilon_{kij} u_i    = \left( \matrix 0 & 0 & 0 & 0 
\cr 0 &0 & -z_0  & y_0  \cr 0 &  z_0 & 0 & -x_0 \cr 0 & 
-y_0 & x_0 & 0 \endmatrix \right) \ .
$$ 
The fibres of $\kappa({\Cal D}_{\sssize \Inv})$ are two 
dimensional and the inclusion into ${\Cal D}_{\sssize 
\Inv}$ is easily found by computing the null space of the 
transpose of the matrix above.

The sections 
$$
S_0 = du_0 \wedge \nu  \quad {\text {\rm and}} \quad
S_1 = \sum _{i=1}^3 x^i du_i \wedge \nu 
$$
of ${\Cal D}$ are invariant. The reduced operator is easily 
obtained {}from
$$
\Delta_{\Inv} = \left( v_{rr} +\frac{2}{r} v_{r}  -r 
w_{tr}- 3 w_t \right) S_0 + \left(  w_{tt} - \frac{1}{r} 
v_{tr}  \right) S_1 \ .
$$
\smallskip
To compute the Palais condition \equationlabel{2}{223}{} we 
have
$$
[Vert_p \kappa(E) ]^{G_x} = \left( \matrix a \cr b x_0 \cr 
b y_0 \cr b z_0 \endmatrix \right)\ ,\  
[Vert^*_p \kappa(E) ]^{G_x} = \left( a, b x_0, b y_0, b z_0 
\right) \ \quad a,b \in \reals 
$$
and the Palais condition \equationlabel{2}{223}{} is 
satisfied. The cohomology condition 
\equationlabel{2}{208}{} is
also easily checked, and the principle of symmetric 
criticality holds. 
(Note that this also follows {}from the symmetry group being 
compact.) 

For completeness we compute the reduced Lagrangian. The 
Lagrangian in \eq{300} restricted to $\Inv^k(E)$ is
$$
\lambda\big|_{\Inv} =  -\frac{1}{2} (r w_t - v_r)^2 \nu \ ,
$$
and evaluating $\lambda \big|_{\Inv}$ on the chain in 
\equationlabel{2}{200}{} we obtain the reduced Lagrangian
$$
\tillambda = \lambda\big|_{\Inv} (\chi) = 
\frac{1}{2} (r w_t - v_r)^2 r^2 dt \wedge dr \ .
$$
The reduced equations agree with the Euler-Lagrange 
equations of the reduced Lagrangian.

\bigskip

\proclaim{Example 3.2} $\Gamma = \{ V_1  =  \partial_y, 
V_2= \partial_t - \partial _z , V_3= (t+z) \partial_y + 
y( \partial_t - \partial _z) + (u_3-u_0) \partial_{u_2} - 
u_2 (\partial_{u_0} + \partial_{u_3})    \} $  
\endproclaim

At the point $ \bx_0 = (t_0,x^k_0)$ the isotropy 
vector-field is
$$
Z = V_3 -(t_0+z_0) V_1 - y_0 V_3 = (u_3 - u_0) 
\partial_{u_2} - u_2 (\partial_{u_0} +\partial_{u_3})
 \in \Gamma_{\bx_0} \ .
\Tag303
$$
The sub-bundle $\kappa(E) \subset E $ at the point $\bx_0$ 
is given by $ u_3 = - u_0\, , \, u_2 =0 $
and the inclusion $\kappa (E) \to E $ is 
$$
\iota (t,x,y,z\, ; \, v, w ) \to (t,x,y,z\, ;\, v, w ,0 , 
v ) \ .
$$
The functions $r = t+z $ and $x$ are invariants and the 
kinematic reduction diagram is
$$
\CD
    (r, x \, ; \, v, w)  @<\q_{\kappa }<< (t,x^i\, ; \,  v 
,w)     @> \iota >>  (t,x^i\, ;  \, u_a)
\\
              @V\tilde \pi VV              @V\pi VV      @ 
VV  \pi V
 \\
              (s,x)           @<\q_M < <                
(t,x^i)            @>> id  >  (t,x^i).
\endCD
\Tag304
$$
The invariant sections are then
$$
u_0 = v(r,x) \quad , \quad u_1 = w(r,x) \quad , \quad  u_2 
= 0 \quad , \quad  u_3 = v(r,x) 
$$
and the reduced operator is 
$$
\tilDelta = \left( v_{xx} - w_{rx} \right) 
(du_{0} - du_{3})\wedge \nu   \ .
$$

The infinitesimal linear isotropy representation is
$$
\rho_{p_0} (Z)  =  \left( \matrix 0 & 0 & -1 & 0 \cr 0 &0 & 
0  & 0  \cr -1 &  0 & 0 & 1 \cr 0 & 0 & -1 & 0 \endmatrix 
\right) \ .
$$

To compute the Palais Condition \equationlabel{2}{223}{} we 
have
$$
[Vert_p \kappa(E) ]^{G_x} = \left( \matrix a \cr b  \cr 0 
\cr a \endmatrix \right)\ ,\
[Vert_p \kappa(E)^* ]^{G_x} = \left( a, b , 0, -a \right) \ 
\quad a,b \in \reals
$$
and
$$
[ Vert^*_p(E) ]^{G_x} \cap \left( [Vert_p(E)]^{G_x} \right) 
^\perp = \{ (a,0,0,-a) \ , \, a \in \reals\} \neq 0 
$$
so the principle of symmetric criticality fails.  In fact 
the Lagrangian \eq{300} restricted to the
invariant sections is easily computed to be
$$
\lambda\big|_{\Inv} = 0 \ .
$$
Note however that the Lie algebra cohomology condition 
\equationlabel{2}{208}{} is satisfied
so that the Palais condition is independent of the 
cohomology condition \equationlabel{2}{208}{}.

\bigskip
\bigskip

\InitializeRef

\heading References \endheading

\end